\pdfoutput=1

\documentclass[preprint,12pt,authoryear]{elsarticle}

\usepackage{amssymb}
\usepackage{amssymb}
\usepackage[utf8]{inputenc}
\usepackage{amsmath}
\usepackage{amsfonts}
\usepackage{mathtools}  
\usepackage{amssymb}
\usepackage{graphicx}
\usepackage[running]{lineno}
\usepackage{setspace}
\usepackage{amsmath,amsthm}

\usepackage{lineno}
\makeatletter
\def\ps@pprintTitle{%
 \let\@oddhead\@empty
 \let\@evenhead\@empty
 \def\@oddfoot{\centerline{\thepage}}%
 \let\@evenfoot\@oddfoot}
\makeatother
\begin{document}

\begin{frontmatter}



\title{Increased Efficiency in the Second-Hand Tire Trade Provides Opportunity for Dengue Control}


\author[label1]{Emilene Pliego Pliego}
\ead{emilene.pliego@alumno.buap.mx}
\author[label1]{Jorge Velázquez-Castro*}
\ead{jorge.velazquezcastro@correo.buap.mx}
\cortext[cor1]{Corresponding author.}
\author[label2]{Markus P. Eichhorn}
\ead{markus.eichhorn@nottingham.ac.uk}
\author[label1]{Andr\'es Fraguela Collar}
\ead{fraguela@fcfm.buap.mx}
\address[label1]{Facultad de Ciencias F\'isico-Matem\'aticas, Benem\'erita Universidad Aut\'onoma de Puebla, Avenida San Claudio y 18 Sur, Col. San Manuel, Puebla, M\'exico}
\address[label2]{School of Life Sciences, University of Nottingham,University Park, Nottingham, NG7 2RD, UK}

\begin{abstract}
Dengue fever is increasing in geographical range, spread by invasion of its vector mosquitoes. The trade in second-hand tires has been implicated as a factor in this process as they act as mobile reservoirs of mosquito eggs and larvae. Regional transportation of tires can create linkages between rural areas with dengue to disease-free urban areas, potentially giving rise to outbreaks even in areas with strong local control measures. In this work we sought to model the dynamics of mosquito transportation via the tire trade, in particular to predict its role in causing unexpected dengue outbreaks through vertical transmission of the virus across generations of mosquitoes. We also aimed to identify strategies for regulating the trade in second-hand tires, improving disease control. We created a mathematical model which captures the dynamics of dengue between rural and urban areas, taking into account the movement, storage time of tires, and mosquito diapause. We simulate a series of scenarios. First a mosquito population is introduced to a dengue-free area via movement of tires, either as single or multiple events, increasing the likelihood of a dengue outbreak. An endemic state can be induced regardless of whether urban conditions for an outbreak are met, and an existing endemic state can be enhanced by vector input. Finally we assess the potential for regulation of tire processing as a means of reducing the transmission of dengue fever using a specific case study from Puerto Rico. Our work demonstrates the importance of the second-hand tire trade in modulating the spread of dengue fever across regions, in particular its role in introducing dengue to disease-free areas. We propose that regulation of tire storage and movement can play a crucial role in containing outbreaks and dengue spread.
\end{abstract}


\end{frontmatter}


\section*{Introduction}

Dengue fever is among the most widespread vector-borne diseases, with approximately 2.5~billion people at risk and 50 million infections annually \cite{OMS}. Dengue is endemic in over 100 tropical and subtropical countries \cite{gublerepidemic2002}. It is also the fastest re-emerging disease \cite{Gordon}, imposing an economic burden alongside the impaired health of affected individuals. Two mosquito species are responsible for transmission of the virus via infective bites. The most common vector is \textit{Aedes aegypti}, but the asian tiger mosquito \textit{Aedes albopictus} is increasingly important due to a rapidly expanding global distribution encompassing most tropical regions ~\cite{belli2015, Giovanni}. There are four dengue virus serotypes \cite{Garcia}, and once an individual has been infected by one serotype they are permanently immune to that serotype but only temporarily immune to the others ~\cite{Garcia,Esteva2003}. 

Second-hand tires are widely traded both locally and globally. In countries with \emph{Ae. aegypti} mosquitoes, these often contain standing rain water and eggs ~\cite{Giovanni,Yee}, providing excellent larval habitats and frequently infected with both species ~\cite{honorio2006,Vietnam}. Tires have been an important dispersal mechanism for both mosquitoes and dengue virus. \emph{Ae. albopictus} originated in Asia but invaded the New World in the 1980s via imported used tires and bamboo plants ~\cite{belli2015,gublerepidemic2002}. It is now present in 20 countries in the Americas ~\cite{belli2015}. International trade in used tires and bamboo has also been implicated in the introduction of \emph{Ae. albopictus} to Europe ~\cite{medlocka2012}.

\textit{Ae. aegypti}  was eliminated in 1960 in almost all the New World. However, subsequent social and economic changes in the Americas have permitted the rapid re-infestation of the vector and dengue virus throughout the region. From 1960 to 1990, the annual production of tires increased from 2 to 17 million ~\cite{briseno-garciapotential1996}, suggesting that the management of their movement, disposal and recycling could be an important modulator of dengue dispersal. For example, between 1981 and 1996, Cuba lacked any dengue transmission. Reintroduction has now occurred in some areas; the municipality of Santiago de Cuba was reinfested in 1992 by {\em Ae. aegypti} transported in tires ~\cite{kouri1998}. 

Two processes play an important role in the transmission of \emph{Aedes} and dengue fever via tires. The first is the diapause phase in the mosquito life cycle, enabling eggs to survive long periods of unfavorable conditions, including desiccation ~\cite{diapause}. Vertical or transovarial transmission of dengue also occurs, with infected females passing the virus to their eggs ~\cite{Esteva2000,Martinez2007,martins2012,Murillo}. Emerging adults are therefore able to transmit the disease without first interacting with an infected host ~\cite{Gluber,Gordon}, potentially causing outbreaks in dengue-free areas.

Here we assess the potential role of transportation of tires containing infected eggs in causing outbreaks in areas otherwise free of both vectors and dengue. Our mathematical model is based on two patches, representing a rural area with endemic dengue and an urban area which begins as dengue-free. We incorporate vertical transmission, diapause during transportation, and the efficiency of tire processing. Through this we generate scenarios in which (a) mosquitoes arrive but with no dengue outbreak, (b) a dengue outbreak occurs, (c) an effective endemic state is created due to continuous influx of infected eggs, and (d) an existing endemic infection is enhanced through additional input of infected vectors. In order to assess the potential for management, we present a case study of implementing a management program to reduce tire processing times. Our work demonstrates that, if effectively regulated, a reduction in the time that tires are stored could aid in dengue control.

\section*{Methods}

Our model aims to capture the dynamics of dengue fever in both humans and female mosquitoes. The total human population $N$ remains constant throughout. The modeled populations are divided into rural and urban sections, with movement occurring only in the reservoirs containing eggs. The systems in the two patches differ mainly by the loss and gain of eggs from rural to urban areas. The total human populations in rural area $N_{R}$ and urban area $N_{U}$ are constant. Table ~\ref{table1} describes the population classes.

\begin{table}[h]
	\begin{tabular}{cp{.7\textwidth}}
		\hline
		\textbf{Population}	& \textbf{Description}\\
		$S_{R}$	&  Susceptible human in rural area.\\
		$I_{R}$	&  Infectious human in rural area.\\
		$R_{R}$	&  Recovered human in rural area.\\
		$S_{U}$	&  Susceptible human in urban area.\\
		$I_{U}$	&  Infectious human in urban area.\\
		$R_{U}$	&  Recovered human in urban area.\\
		$M_{SR}$&  Susceptible mosquitoes in rural area.\\
		$M_{IR}$&  Infectious mosquitoes in rural area.\\
		$E_{SR}$&  Susceptible eggs in rural area.\\
		$E_{IR}$&  Infectious eggs in rural area.\\
		$M_{SU}$&  Susceptible mosquitoes in urban area.\\
		$M_{IU}$&  Infectious mosquitoes in urban area.\\
		$E_{SU}$&  Susceptible eggs in urban area.\\
		$E_{IU}$&  Infectious eggs in urban area.\\
		\hline	
	\end{tabular}\caption{State variables.}\label{table1}\end{table}

The susceptible human class has a per-capita birth rate $\eta$, and an identical per-capita death rate $\eta$, meaning that the overall population remains stable through time. Individuals in this class become infectious according to the bite rate of infected vectors $\alpha$. The vectors become infected by biting infectious hosts with an identical contact rate $\alpha$. The rate at which humans recover from infection, whereupon they become permanently immune, is $\gamma$.

In contrast, mosquitoes never recover from the disease. Our model only considers the fraction $\kappa$ of mosquitoes that are female, as males do not transmit the disease.
Mosquito populations increase through egg eclosion according to the development rate $\omega$ and individuals die with a rate $\epsilon$. Female mosquitoes oviposit at a rate $\phi$ and the eggs have an intrinsic mortality rate $\pi$. If a female mosquito is already infected, a fraction $\nu$ of its oviposited eggs are infected (vertical transmission). The hatcheries have a carrying capacity $C_{a}$ where $a\in\{r,u\} $, and $r$ is rural and $u$ urban area.

\begin{figure}[!h]
	\centering
	\includegraphics[scale=.5]{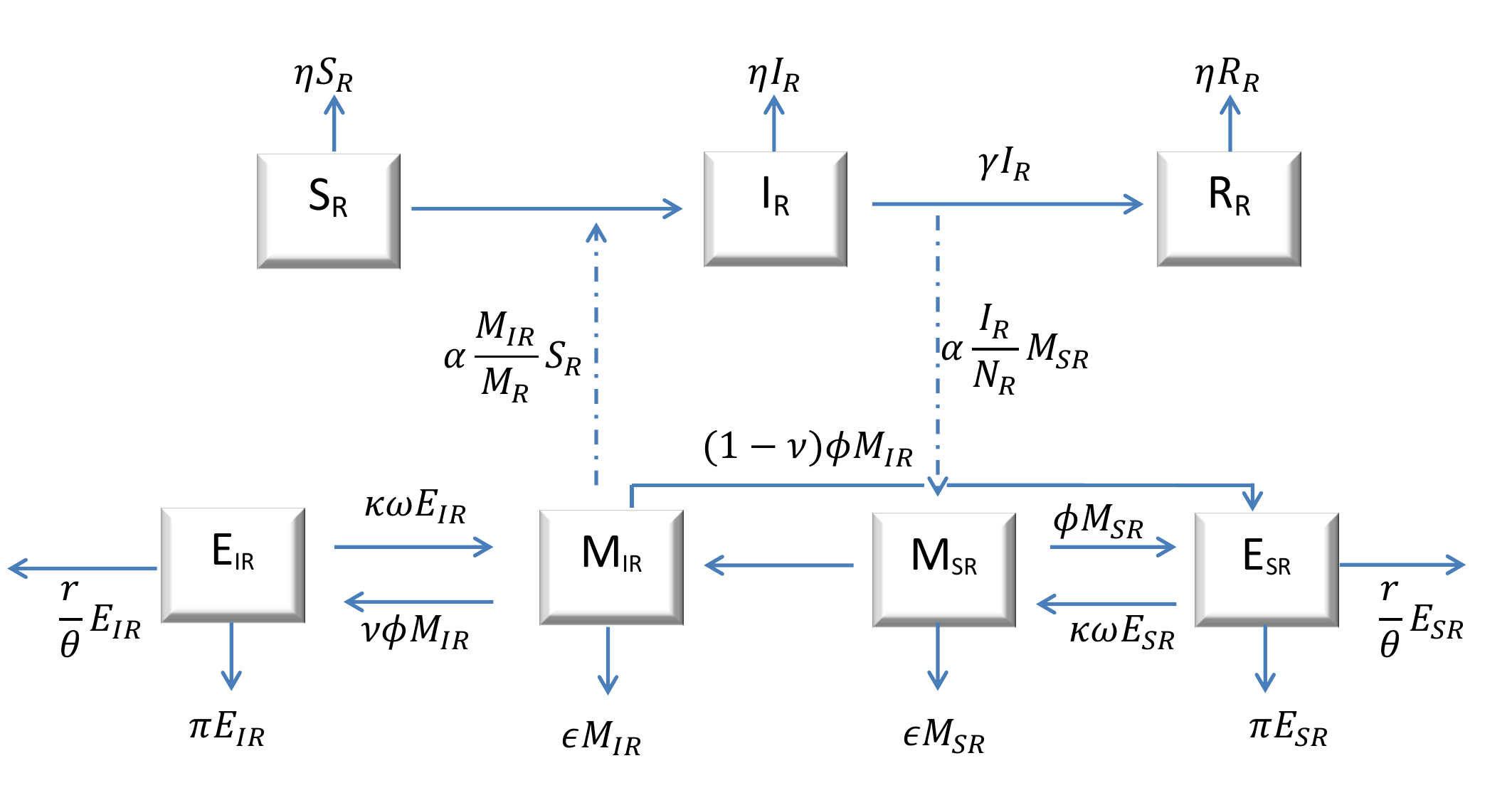}
	\caption{{ {\bf Flowchart from rural dengue fever model.} Elements of the upper row refer to segments of the human population, whereas the lower row refers to adult mosquitoes (M) or their eggs (E). See Tables 1 and 2 for definitions of terms. }}
	\label{fig1}
\end{figure}

The number of tires transported from rural area to urban area per unit time is $r$ and $\theta$ is the mean number of tires in the rural area.
Hence $r E_{IR}/\theta$ is the rate of egg movement. During transportation a fraction $\chi$ of eggs survive. $\psi(\tau_{s}/\tau_{d})$ is the fraction of eggs in the tires that were able to hatch as adults before being killed by the recycling process of the tire. Thus it is a function of the storage time $\tau_{s}$ before tire processing and the egg development time $\tau_{d}$. $\psi$ should be a function such that when $\tau_{s}=0$ then $\psi = 0$, while when $\tau_{s}$ is greater than the development time of the eggs $\tau_{s}$, $\psi$ should approach one. A flow diagram of the model system is shown in Figs~\ref{fig1}~and~\ref{fig2}. Table~\ref{table2} shows a summary of parameters.

\begin{table}[h]
	\centering\resizebox{14 cm}{!} {
		\begin{tabular}{clcc}
			\hline
			\textbf{Parameter}	&	\textbf{Description}\\
			$\eta$ & Per-capita birth and natural mortality  rates in humans  \\ 
			$\gamma$& Per-capita recovery rate\\
			$\alpha$& Effective biting rate, per day\\
			$C_{a}$ &  Carrying capacity of hatcheries, where $a\in\{r,u\} $, and $r$ is rural and $u$ urban area\\
			$\phi$ & Number of eggs laid per day for every female mosquito \\
			$\epsilon$ & Per-capita mortality rate of adult mosquitoes \\
			$\pi$ & Per-capita mortality rate of immature stage mosquitoes \\
			$\nu$ & Proportion of eggs that are infected by vertical transmission\\
			$\omega$ & Development rate of immature to mature stages \\
			$\kappa$ & Fraction of mosquitoes that are female\\
			$\frac{r}{\theta}$ & Per-capita egg transportation rate\\
			$\chi$& Fraction of eggs that survive the transportation\\
			$\psi(\tau_{s}/\tau_{d})$ & Fraction of eggs in tires that were able to continue their development before tire processing\\
			$\tau_{s}$,$\tau_{d}$ & Tires storage time and egg development time\\
			\hline
		\end{tabular}}
		\caption{{Model parameters}}\label{table2}
	\end{table}

Following the assumptions and parameters defined above, the system of differential equations that model the dynamics of dengue in human and mosquito populations in the rural area is given by:
	
\begin{eqnarray*}
	\dot{S_{R}}	&=&	 \eta N_{R}- \alpha\frac{S_{R}}{N_{R}}M_{IR}-\eta S_{R},\\
	\dot{I_{R}}	&=&\alpha\frac{S_{R}}{N_{R}}M_{IR}- (\eta  +\gamma)I_{R}, \\
	\dot{R_{R}}&=&	 \gamma I_{R} - \eta R_{R},\\
	\dot{M_{SR}}&=& \kappa\omega E_{SR}- \alpha\frac{I_{R}}{N_{R}}M_{SR} - \epsilon M_{SR},\\
	\dot{M_{IR}}	&=&  \kappa\omega E_{IR}+ \alpha\frac{I_{R}}{N_{R}}M_{SR} - \epsilon M_{IR},\\
	\dot{E_{SR}}	&=& \phi M_{SR}\left(1-\frac{E_{R}}{C_{r}}\right)+ (1-\nu)\phi M_{IR}\left(1-\frac{E_{R}}{C_{r}}\right)- (\pi + \omega +\frac{r}{\theta})E_{SR},\\
	\dot{E_{IR}}	&=& \nu\phi M_{IR}\left(1-\frac{E_{R}}{C_{r}}\right)- (\pi + \omega + \frac{r}{\theta}) E_{IR}.
\end{eqnarray*}
	
The differential equations that model the dynamics of dengue disease in human and mosquito populations in the urban area are given by:
	
\begin{figure}[!h]
	\centering
	\includegraphics[scale=.5]{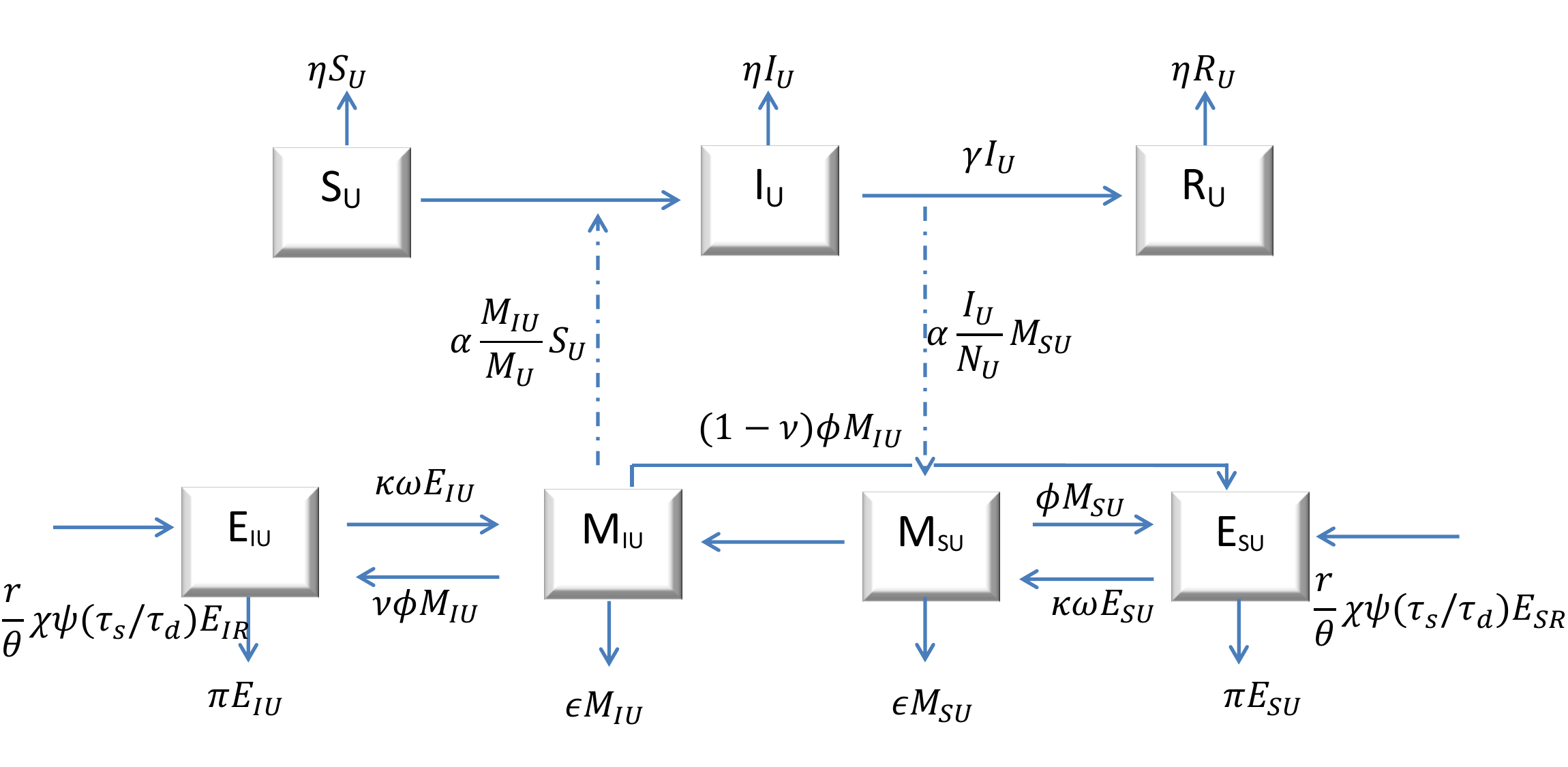}
	\caption{{ {\bf Flowchart from urban dengue fever model.} Elements of the upper row refer to segments of the human population, whereas the lower row refers to adult mosquitoes (M) or their eggs (E). See Tables 1 and 2 for definitions of terms.}}
	\label{fig2}
\end{figure}

\begin{eqnarray*}
	\dot{S_{U}}	&=&	 \eta N_{U}- \alpha\frac{S_{U}}{N_{U}}M_{IU}-\eta S_{U},\\		\dot{I_{U}}	&=&\alpha\frac{S_{U}}{N_{U}}M_{IU}- (\eta  +\gamma)I_{U},\\
	\dot{R_{U}}&=&	 \gamma I_{U} - \eta R_{U},\\
	\dot{M_{SU}}&=& \kappa\omega E_{SU}- \alpha\frac{I_{U}}{N_{U}}M_{SU} - \epsilon M_{SU},\\
	\dot{M_{IU}}	&=&  \kappa\omega E_{IU}+ \alpha\frac{I_{U}}{N_{U}}M_{SU} - \epsilon M_{IU},\\
	\dot{E_{SU}}	&=& \phi M_{SU}\left(1-\frac{E_{U}}{C_{u}}\right)+ (1-\nu)\phi M_{IU}\left(1-\frac{E_{U}}{C_{u}}\right)- (\pi + \omega)E_{SU}+...\\
		&&\frac{r}{\theta}\chi\psi\left(\frac{\tau_{s}}{\tau_{d}}\right)E_{SR},\\
	\dot{E_{IU}}	&=& \nu\phi M_{IU}\left(1-\frac{E_{U}}{C_{u}}\right)- (\pi + \omega) E_{IU}+\frac{r}{\theta}\chi\psi\left(\frac{\tau_{s}}{\tau_{d}}\right)E_{IR}.
\end{eqnarray*}
	
Where
\begin{eqnarray*}
	N_{R}= S_{R}+I_{R}+R_{R}, \\
	N_{U}= S_{U}+I_{U}+R_{U}, \\
	M_{R}=M_{SR}+M_{IR}, \\
	E_{R}= E_{SR}+E_{IR}, \\
	M_{U}=M_{SU}+M_{IU}, \\
	E_{U}=E_{SU}+E_{IU}.
\end{eqnarray*}
	
The model explicitly takes into account the movement and storage time of tires. Our study focuses on the necessary conditions for four possible outcomes. Scenario $I$ considers the introduction to the urban area of mosquitoes from a rural area. In Scenario $II$ a dengue outbreak emerges in the urban area as a consequence of the joint introduction of the mosquito and the virus in infected eggs. Scenario $III$ induces or enhances an endemic state in the urban area through the constant introduction of infected mosquitoes eggs. Finally, in Scenario $IV$, we consider how regulation of the market in second hand tires could act as a dengue control measure. To demonstrate the impacts of dengue spread we calculate the secondary dengue cases generated in the urban area as the result of a single case in rural area. This quantity can be used as a preliminary measure of the impact of controlling the movement of tires during dengue outbreaks. Finally we apply our model to a specific study of the tire management system in Puerto Rico using data from the Solid Waste Authority (ADS) ~\cite{PuertoRico}.

\section*{Results}
	
\subsection*{Introduction of mosquitoes to an urban area} 
	
Transportation of a single batch of tires can led to the introduction of a mosquito population from rural to urban areas. In order to obtain the conditions when this might occur, we determine the \textit{urban offspring reproduction number} (derived in \ref{Appendix A}) $R^{u}_{M}$  by means of the next generation matrix ~\cite{Brauer2008}.
If $R_{M}^{u}>1$, then the population of mosquitoes is able to establish itself from a small number of eggs, while if the environmental conditions cause $R_{M}^{u}<1$ the mosquito population will eventually become extinct. In a single batch of $N_{T}$ tires, the number of viable eggs that arrive in the urban area is given by
\[\frac{\omega}{\pi+\omega}\psi\left(\frac{\tau_{s}}{\tau_{d}}\right)\chi N_{T}\frac{E^{*}_{R}}{\theta}=\frac{\omega}{\pi+\omega}\psi\left(\frac{\tau_{s}}{\tau_{d}}\right)\chi N_{T}\frac{R_{M}^{r}-1}{R_{M}^{r}}\frac{C_{r}}{\theta}\] where $\frac{\omega}{\pi+\omega}$ is the probability of an egg hatching into an adult mosquito, $E^{*}_{R}$ is the stationary number of eggs in the rural area, $R_{M}^{r}$ is the \textit{rural offspring reproduction number} (see \ref{Appendix B}), and $\chi\psi(\frac{\tau_{s}}{\tau_{d}})$ is the fraction of eggs that survive before the tire processing cycle completes. 
	
The introduction of a mosquito species will occur when the following conditions are met:

\begin{eqnarray}
	\label{cond_specie}
	R^{u}_{M}=\frac{\kappa\omega\phi}{\epsilon(\pi+\omega)}&>&1 \nonumber \\
	&&\\
	{\rm and}\qquad \frac{\omega}{\pi+\omega}\frac{N_{T}}{\theta}\chi\psi\left(\frac{\tau_{s}}{\tau_{d}}\right)\frac{R_{M}^{r}-1}{R_{M}^{r}}C_{r} & > & 1\nonumber 
\end{eqnarray}
	
The first condition indicates that a mosquito population is sustainable in the urban area, while the second condition represents the transportation of more than one successful egg. 
	
If tire recycling becomes an established market with a constant flux of tires from the rural to the urban area, then the expected waiting time $T_{M}$ before the introduction of a mosquito species from the rural to the urban population is given by the inverse of the rate of egg introduction:

\begin{equation}\label{Tm}
	T_{M}=\left[\frac{\omega}{\pi+\omega}r\chi\psi\left(\frac{\tau_{s}}{\tau_{d}}\right)\frac{R_{M}^{r}-1}{R_{M}^{r}}\frac{C_{r}}{\theta} \right]^{-1} \qquad {\rm for} \qquad R_{M}^{r}>1
\end{equation}
	
\subsection*{A dengue outbreak occurs}
	
If continuous introduction of tires takes place from a dengue-endemic rural area to the urban area, a dengue outbreak might be precipitated by transportation of infected eggs. In order for this to happen the conditions (~\ref{cond_specie}) must be met, but also requires that the \textit{basic reproductive number without vertical transmission} in the urban area $R^{u}_{0}>1$. Its value is given by $R_{0}^{u}=\sqrt{\frac{\alpha}{\epsilon}\frac{\beta N}{(\eta + \gamma)M^{*}}}$ (derived in \ref{Appendix C}). In this we omit vertical transmission as its effect is negligible at the beginning of an outbreak. Thus:
	
\begin{equation}\label{cond_Ro}
	\sqrt{\frac{\alpha}{\epsilon}\frac{\beta N}{(\eta + \gamma)M^{*}}} > 1 
\end{equation}
should be meet in addition to (~\ref{cond_specie}).
	
The characteristic waiting time before an outbreak $T_{o}$ is given by the inverse of the effective rate of introduction of infected female mosquitoes:
	
\begin{equation}\label{waiting_time}
	T_{o}=\left[ \frac{\kappa\omega}{\pi+\omega}r\chi\psi\left(\frac{\tau_{s}}{\tau_{d}}\right)\frac{E^{*}_{IR}}{\theta} \right]^{-1} 
\end{equation}
where $\frac{r}{\theta}\chi\psi\left(\frac{\tau_{s}}{\tau_{d}}\right)$ is the fraction of successfully imported eggs, $\frac{\kappa\omega}{\pi+\omega}$ is the probability of an egg hatching into a female mosquito before dying of natural causes. 
	
In the case of introduction of a single batch of $N_{T}$ tires, in addition to (\ref{cond_specie}) and (~\ref{cond_Ro}), the following condition must be satisfied:
	
\begin{equation} \label{condicion}
	\frac{\kappa\omega}{\pi+\omega}N_{T}\chi\psi\left(\frac{\tau_{s}}{\tau_{d}}\right)\frac{E^{*}_{IR}}{\theta} > 1
\end{equation}
	
This effectively states that more than one infected female mosquito needs to be introduced.

\subsection*{Endemic dengue states can be induced and enhanced}
	
\begin{figure}[!h]
	\centering
	\includegraphics[scale=.25]{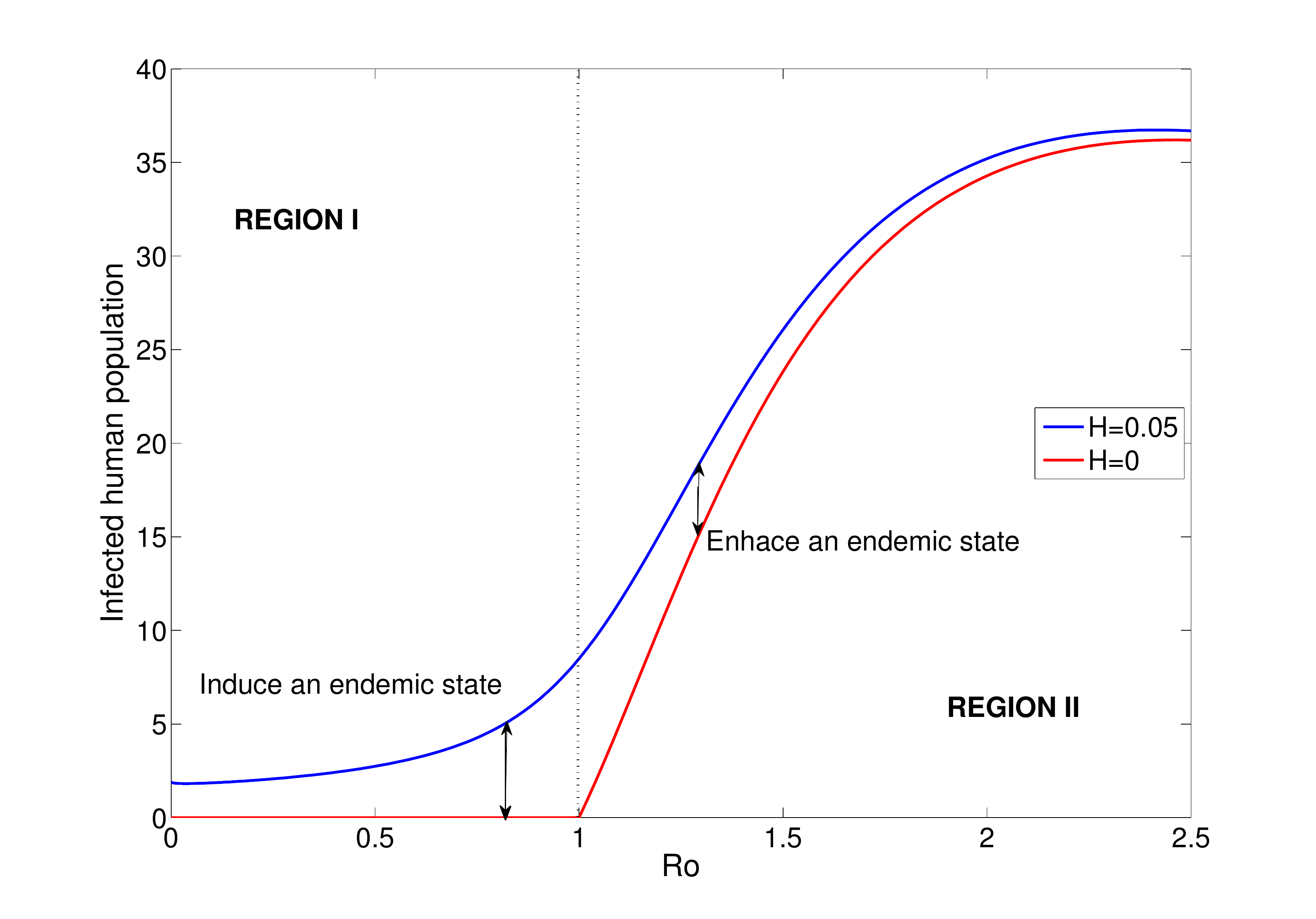}
	\caption{{\bf Enhance an endemic state.} It is possible to induce a dengue-endemic state even though $R_{0}^{u}<1$ (Region I) if there is  a continuous flow of tires from an endemic rural area. If the disease is already endemic, tire transport of eggs will enhance the endemic state (Region II). $H=\frac{r}{\theta}\chi\psi\left(\frac{\tau_{s}}{\tau_{d}}\right)$ represents variation of the flow parameters of tires.}
	\label{fig3}
\end{figure}

\begin{figure}[!h]
	\centering
	\includegraphics[scale=.25]{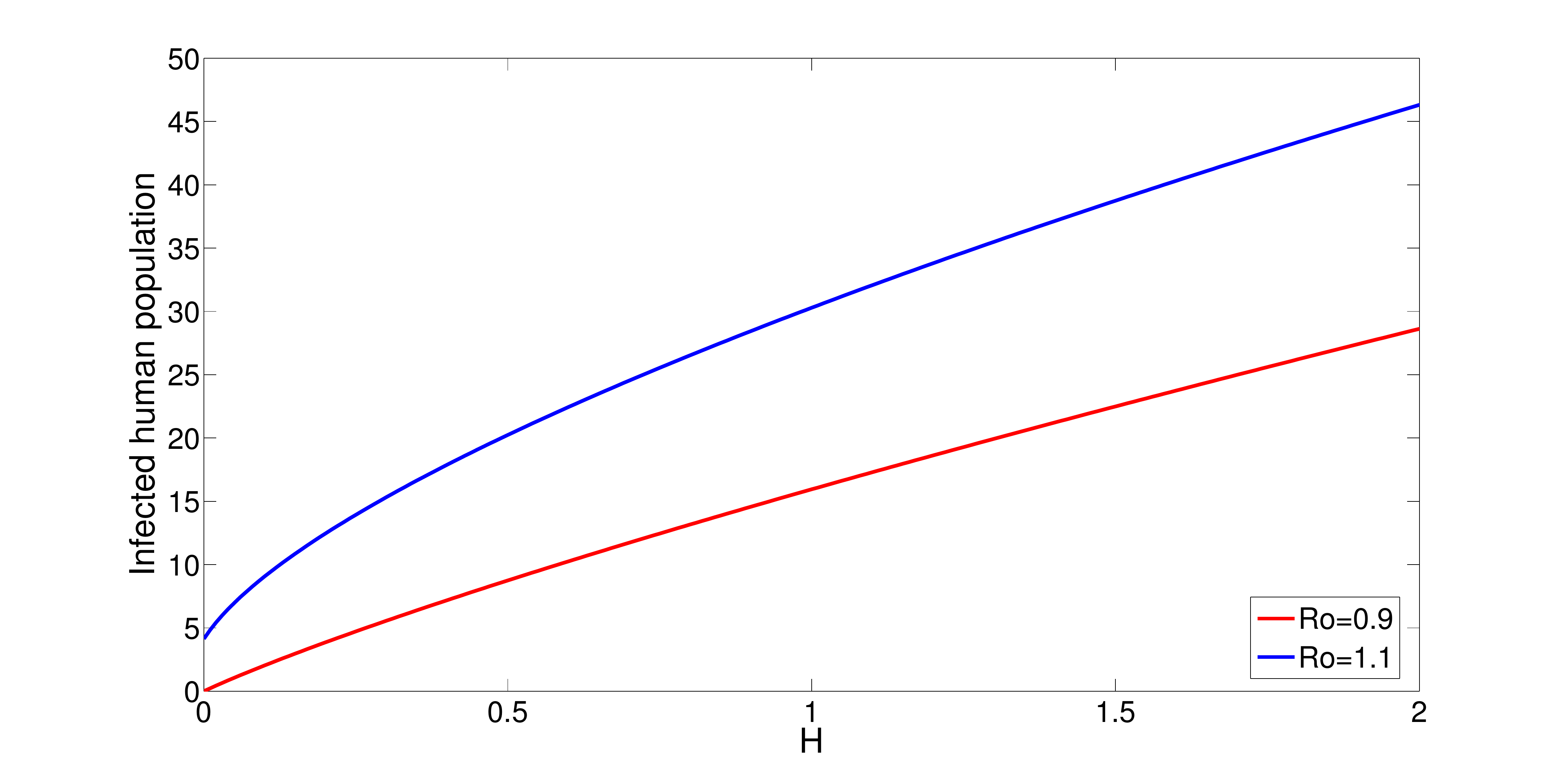}
	\caption{{\bf Stationary level of infection in the urban population with increasing introduction of eggs.} Dengue cases increase as the number of introduced infected eggs per unit time $H$ is increased. Red line $R_{0}^{u}<1$ (intrinsically non-endemic state) and blue line $R_{0}^{u}>1$ (endemic state).}
	\label{fig4}
\end{figure}

There may be situations in which $R_{0}^{u}<1$, and therefore dengue infestation is not self-sustaining, but where continuous introduction of infected eggs in tires from an endemic rural area can induce an endemic state in the urban area. This state is not maintained by the intrinsic dynamics of the disease in the urban area and will cease if the introduction of infected eggs is interrupted see Figs ~\ref{fig3} and ~\ref{fig4}.
	
In this situation, the expected number of dengue cases at any given time $I^{*}_{U}$ is determined as
	
\begin{equation}
	I^{*}_{U}=\frac{M^{*}_{IU}}{M^{*}_{IU}-N_{U}\frac{\eta}{\alpha}} \label{estationary_Infected}
\end{equation}
where $M^{*}_{IU}$ is found from the solution of the following quadratic equation
	
\begin{align}
	(R_{o}^{u})^{2}(\eta + \gamma)(\frac{\eta + \gamma}{\beta} + N_{u}) (M^{*}_{IU})^{2} & \nonumber \\   
	- (\eta + \gamma)\left((R_{o}^{u})^{2}(\frac{\kappa\omega}{\beta}E_{IU}^{*} + N_{U}M_{U} \right. & )-  \eta N_{U}\bigg)M^{*}_{IU}  \label{m_polinom} \\
	&-\frac{\kappa\omega}{\epsilon}\eta N_{U}E_{IU}^{*} = 0 \nonumber
\end{align}
and $E^{*}_{IU}$ is given by
	
\begin{equation*}
	E_{IU}^{*}=\frac{r\chi\psi(\frac{\tau_{s}}{\tau_{d}})}{\theta(\omega + \pi)}E^{*}_{IR}
\end{equation*}

Where dengue is already endemic in the urban area, the continuous importation of tires can enhance the number of infected people \textit{see Figs ~\ref{fig3} and ~\ref{fig4}}.  The number of infections at any given time is given by (\ref{estationary_Infected}) when $R_{0}^{u}>1$ in (\ref{m_polinom}).

\subsection*{Regulation of the second hand tire market as a dengue control measure}
	
If tires are processed immediately, or at least soon after arrival to the urban area such that $\tau_{s}<<\tau_{d}$, introduction of dengue fever does not occur. This is because eggs die during tire processing before hatching.

In the rural area, diminishing the number of eggs reduces \textit{rural reproduction number (see \ref{Appendix D})} $R_{0}^{r}$ and hence the number of dengue cases. In order for this to occur, the reduction of eggs in the rural area should change the abundance of offspring from $R_{M}^{r}>1$ to $R_{M}^{r}<1$. If this does not happen then the risk of introducing infected eggs into the urban area will persist unless tires are processed immediately. The maximum storage time $\tau_{s}$ with which to minimize the risk of dengue introduction can be estimated by ensuring that the expected latency before introduction of an infected female mosquito (~\ref{waiting_time}) is greater than the extinction time of the vector in the rural area.
	
We can use the Jacobian matrix when $R^{r}_{M}<1$ of the vector demography (~\ref{demography}), to obtain the characteristic extinction time. 
The inverse of the smallest absolute value from its eigenvalues is an estimator of its extinction time. Thus, the following condition should be met to reduce the risk of dengue dispersal in an established market where tires are continuously imported to the urban area:
	
\begin{equation*}
	T_{o} < \left|\frac{1}{2}(\gamma+\sqrt{\xi})\right|^{-1} \quad {\rm and} \quad R_{M}^{r}=\frac{\kappa\omega\phi}{\epsilon(\pi+\omega + r/\theta)}<1
\end{equation*}
where $\gamma=-(\epsilon+\pi+\omega+r/\theta)$, $\xi=\gamma^{2}-4\Xi$ and $\Xi=\epsilon(\pi +\omega + r/\theta)(1- R^{r}_{M})$ (see \ref{Appendix B}). This simultaneously works as a control measure in the rural area.

In order to assess the impact of interventions in the tire trade on disease dynamics, we can calculate the secondary human infections in the urban disease free area caused by human infections in the rural area at the beginning of an outbreak $R_{r\rightarrow u}$ (see \ref{Appendix F}). There will be one initial case of dengue virus in the urban area related to tire transportation for each $1/R_{r\rightarrow u}$ cases in the rural area, where
	
\begin{eqnarray*}
		R_{r\rightarrow u}&=&\frac{\alpha\kappa\omega r\chi }{\epsilon(\omega+\pi)(\theta(\omega+\pi)+r)}\psi\left(\frac{\tau_{s}}{\tau_{d}}\right)\frac{\nu\phi}{\epsilon}\frac{\beta}{(\eta+\gamma)}.
\end{eqnarray*}
	
Thus, $R_{r\rightarrow u}$ gives the number of cases in urban area which are derived from an infected person in the rural area.
	
\subsection*{Case study}
	
One of the main barriers for dengue eradication in Latin America is the problem of stored tires. These are a favorable site for the breeding of multiple vector insects, with implications for disease transmission and human health. As mentioned previously, used tires are one of the favored sites in which \textit{Ae. aegypti} females deposit their eggs, becoming an important pathway for their proliferation and thus causing outbreaks of dengue in tropical and subtropical countries. An endemic disease state can arise in areas where the environmental conditions are suitable ~\cite{Tires}.
	
Some Latin American countries, such as Costa Rica and Peru, have banned the import of used tires. This is due to the perceived danger to public health, in addition to concerns regarding road safety and protection of the environment. Costa Rica does not possess the necessary technology to treat used tires without causing environmental pollution ~\cite{Tires}.
	
In Puerto Rico the accumulation of discarded tires in {\em gomeras} and facilities around the island represent an environmental and health crisis. The country has therefore implemented a tire management program. According the Solid Waste Authority (ADS), around $18,000$  tires are discarded every day; this amounts to to $4.7$ million tires a year. Despite the tire management program, it is not possible to recover all discarded tires. Among the major public health risks of excessive accumulation of tires is the spread of pests and diseases such as dengue. A total of 6,766 confirmed cases of dengue were reported in Puerto Rico in 2013. For this reason the authorities have decided to reduce the disposal of tires. They have introduced authorized solid waste facilities in which the accumulation of tires is permitted for up to 90 days. The law also allows local governments to collect used tires voluntarily and temporarily. The collection and transport of discarded tires is carried out by official vehicles ~\cite{PuertoRico}.
	
\begin{figure}[!h]
  \centering
  \includegraphics[scale=.5]{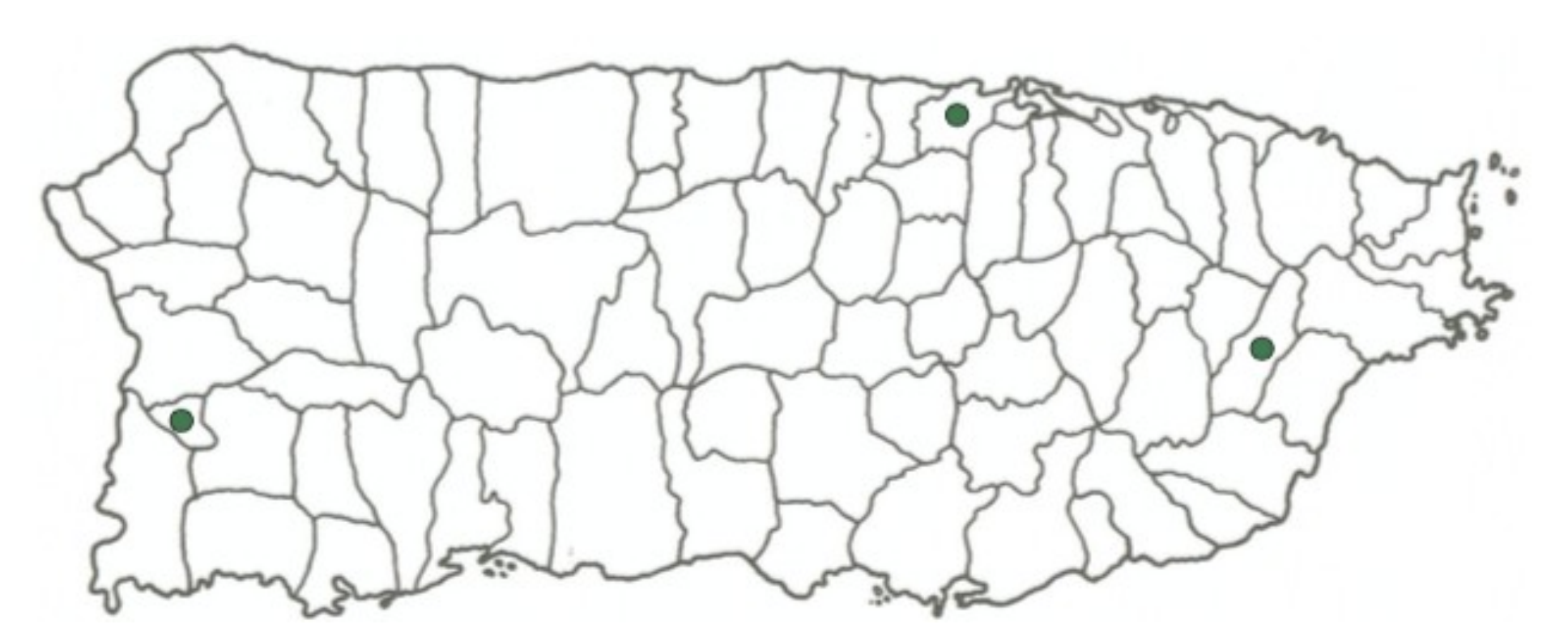}
  \caption{{The Solid Waste Authority (ADS) has detected all the centers of sale and storage of discarded tires (\emph{gomeras}) in each state on the island. The green points show the official temporary storage centers of tires in the municipalities of Hormigueros, Toa Baja and Las Piedras). {\bf Source}: ADS.}}
	\label{fig5}
\end{figure}

Fig~\ref{fig5} shows locations of tire storage centers in Puerto Rico. In this section we employ our model to analyze the implications of tire management for dengue transmission in this specific geographical context.

\begin{table}[h]
	\centering\resizebox{8 cm}{!} {
		\begin{tabular}{cccc}
			\hline
			\textbf{Parameter}	
			& \textbf{Value}&\textbf{Units} & \textbf{Reference}\\
				\hline
				$\eta$  &  $0.002$ &1/days& Estimated \\	
				$\gamma$&$1/7$&1/days&~\cite{adams2010}\\
				$\alpha$& $0.67$&1/days&~\cite{adams2010}\\
				$C_{a}$ & 10000 and 1000 & eggs & Estimated \\
				$\phi$ &$10$&1/days&~\cite{Esteva2006}\\
				$\epsilon$  &$1/8$&1/days&~\cite{adams2010}\\
				$\pi$  &$1/8$&1/days&~\cite{adams2010}\\
				$\nu$ &$0.3$ &proportion&~\cite{adams2010} \\
				$\omega$  &$1/8$&1/days&~\cite{adams2010}\\
				$\kappa$ &$0.5$&proportion& Estimated\\
				$\frac{r}{\theta}$  & $10, 20$&1/days& Estimated\\
				$\chi$&$0.64$ &1/days& Estimated using ~\cite{PuertoRico}\\
				$\psi(\tau_{s}/\tau_{d})$&&proportion\\
				$\tau_{s}$,$\tau_{d}$ & 90, 10&days& ~\cite{PuertoRico},~\cite{Esteva2006}\\
			\hline
		\end{tabular}}
		\caption{{Parameter values for simulating the dynamics of dengue transmission in Puerto Rico based on literature sources.}}\label{table3}
\end{table}
		
\begin{figure}[!h]
	\centering
	\includegraphics[scale=.2]{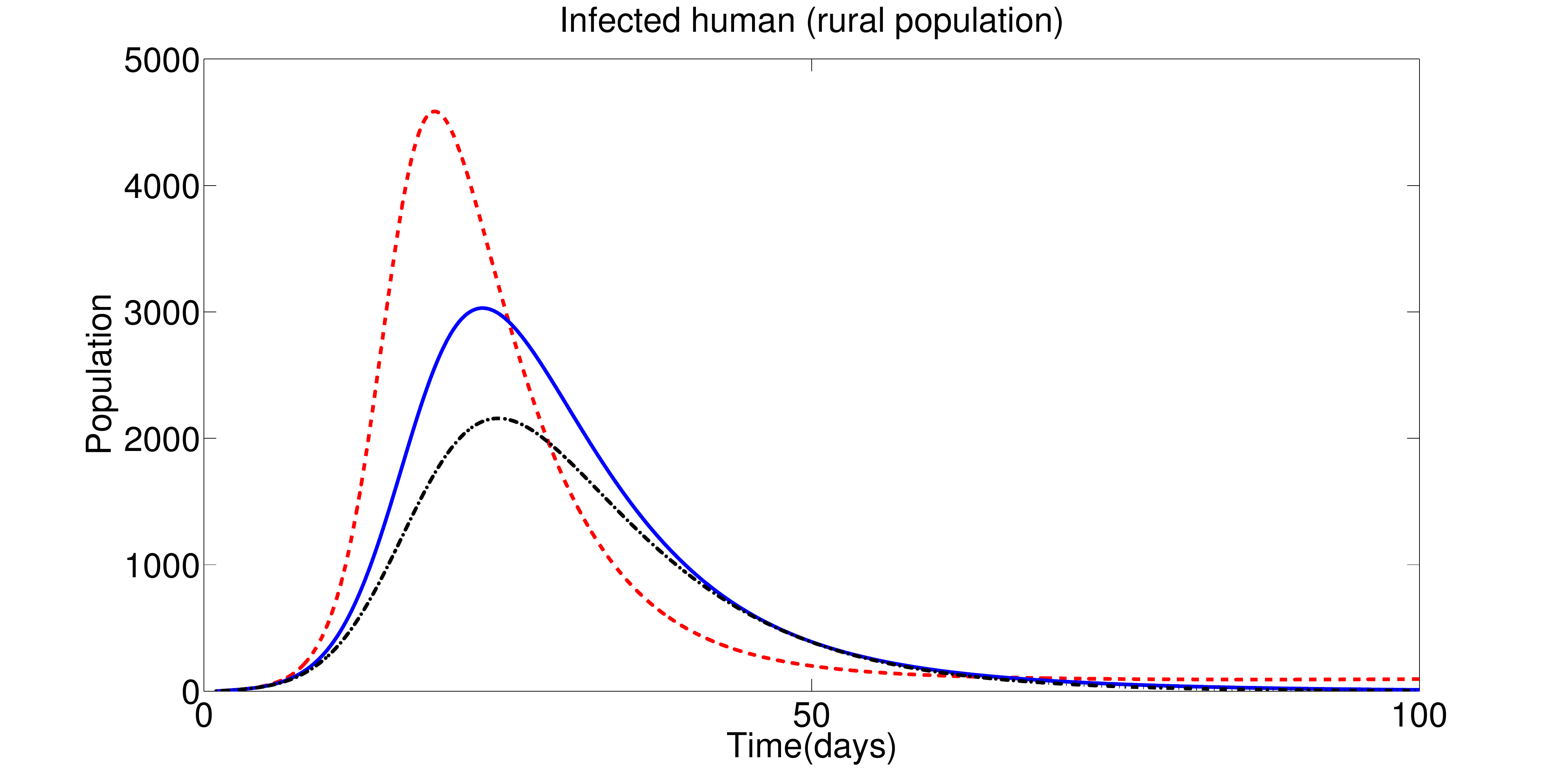}
	\caption{{{\bf Dynamics of dengue infection in rural areas under different scenarios.} Red dashed line shows the population size of infected humans when there is no transportation of tires with infected eggs. Blue solid line shows the infected population when tire transportation takes place, and black dash-dot line shows the infected population when tire transportation occurs  when the average of infected eggs per tire is increased}}
	\label{fig6}
\end{figure}

\begin{figure}[!h]
	\centering
	\includegraphics[scale=.2]{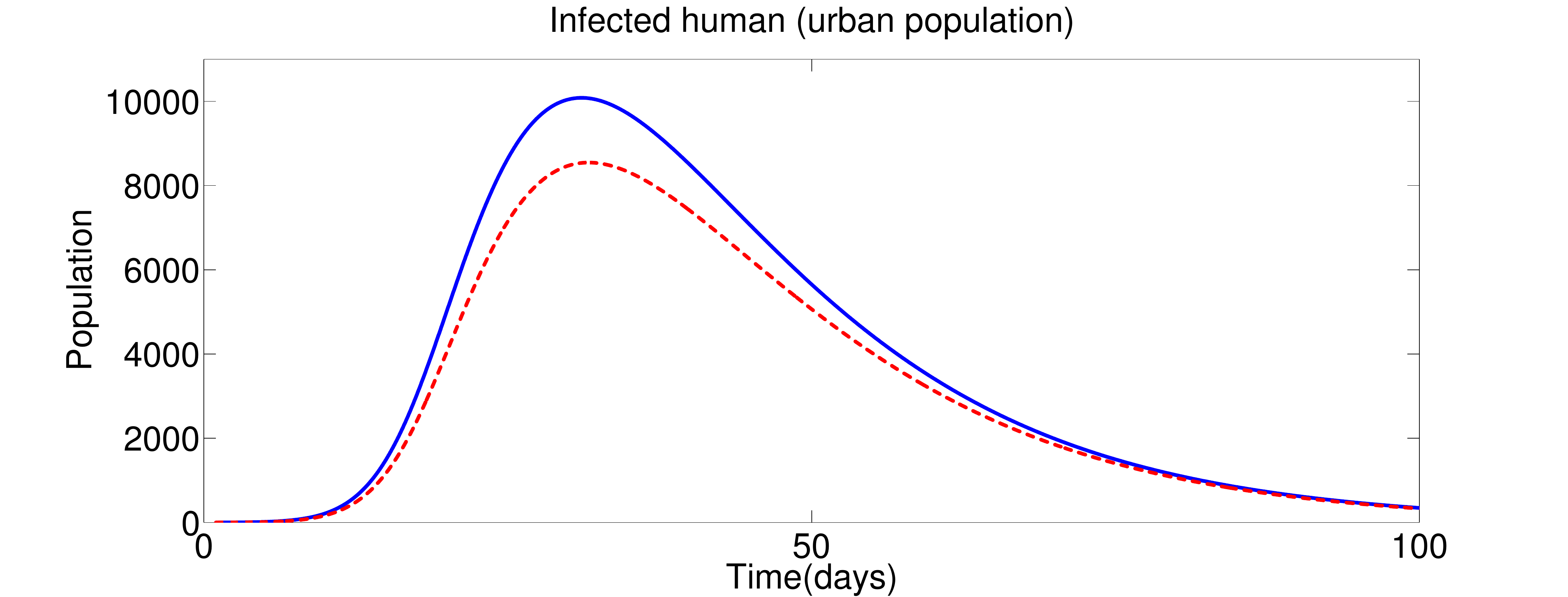}
	\caption{{ {\bf Dynamics of dengue infection in urban areas, indicating the impact of the tire recycling program.} Blue solid line shows the outbreak dynamics with unregulated transport of tires, red dashed line shows the predicted urban outbreak given the existence of this program.}}
	\label{fig7}
\end{figure}

Figs~\ref{fig6}~and ~\ref{fig7} show the populations of infected humans in rural and urban areas. In Fig ~\ref{fig6} the infected population in in rural areas with no transportation of tires {\em (red line)} demonstrates that the disease is endemic. If transportation of tires takes place, the infected population in rural areas declines by $9.79\%$ {\em (blue solid line)} and it becomes possible for the disease to be eradicated. 	

In contrast, if there are around twenty infected eggs per tire and transportation to the storage sites is made systematically, the number of infected people in the rural area is reduced by $28.61\%$ (black dash-dot line). In this situation tire transportation acts as a control measure in rural areas.

Puerto Rico has instituted a program to recover used tires. Fig ~(\ref{fig7}) shows the comparison between number of infected people depending on whether the tires are handled appropriately. The blue line depicts the infected population in the absence of a recycling program. The recycling program recovers and processes $36\%$ of tires in the three temporary storage centers. Taking this into account, the program has reduced the number of infected people in urban areas by $13.04\%$ {\em (red line)}. 

\section*{Discussion}
		
The trade in second-hand tires, if unregulated, is an important factor in the generation of dengue outbreaks and can cause endemic states to be hard to eliminate. Discarded tires are believed to be one of the most productive hatcheries of the \emph{Aedes} mosquitoes which transmit dengue fever \cite{honorio2006}, with many eggs being transported whilst in a diapause state. As a result of vertical transmission across generations, infected mosquitoes can pass the virus to their offspring, and therefore a new generation of infected vectors will emerge able to transmit the disease without first feeding on an infected individual. Evidence also exists of vertical transmission of Zika and Chikungunya viruses by \textit{Ae. aegypti} and \textit{Ae. albopictus} \cite{Thangamani_Vertical_2016,Ferreira-de-Brito_First_2016,Niyas_Molecular_2010,Agarwal_Evidence_2014}, extending the implications of our model to other vector-borne diseases. Our model therefore has important consequence for the spread of a range of emerging diseases. Even in hostile environmental conditions, the resistance of \emph{Aedes} eggs to desiccation, combined with vertical transmission of the virus, is likely to facilitate the persistence and spread of disease. 
		
Our model demonstrates that the movement of tires containing mosquito eggs has the potential to transfer both vector and virus from rural to urban regions, and with a sufficient rate of input, can induce an endemic dengue state in the urban area even if control measures would otherwise cause it to be eradicated. Management of the tire trade to reduce their storage time is a potential strategy for reducing dispersal of the disease, and we demonstrate using an empirical case study from Puerto Rico that even a modest program of tire collection from urban areas can lead to major declines in the disease burden experienced in urban areas. 
		
Tires left in the open are productive \textit{Ae. aegypti} hatcheries ~\cite{Vietnam,yeetires2008,honorio2006}, increasing the risk of dengue transmission. There is also substantial evidence that the transportation of second-hand tires between urban areas has led to the introduction or re-emergence of dengue in areas previously free of disease \cite{belli2015,medlocka2012,kouri1998}. 
		
Our model explicitly takes into account the movement and storage of second-hand tires, typically from rural to urban areas for processing, a common feature of their trade. 
In our study we demonstrate the conditions leading to four scenarios. In the first, the transport of a batch of tires to a zone free of disease introduces the vector species. This occurs when appropriate environmental conditions mean $R_{M}^{u}>1$, and the latency before a mosquito invasion occurs can be estimated as $T_{m}$ (~\ref{Tm}).
Furthermore, the analysis shows that a continuous flow of tires from the endemic rural area, combined with vertical transmission of the virus, can lead to a dengue outbreak in the urban area. This happens when $R_{0}^{u}>1$ (~\ref{cond_Ro}). From this it is possible to estimate the critical size (~\ref{condicion}) of a batch of tires that will trigger an outbreak in the urban area. Moreover, the expected latency before an outbreak $T_{0}$ if unregulated importation of tires takes place can be \mbox{estimated (\ref{waiting_time}).}
		
A third scenario is constructed when dengue is endemic in the rural area. A continuous introduction of infected eggs in tires to the urban area can induce an urban endemic state even though $R_{0}^{u}<1$. This state is not maintained by the intrinsic dynamics of the disease in the urban area, and therefore would abate if the transportation of tires were to be interrupted. On the other hand, if the urban area is already a dengue endemic region, the introduction of tires with infected eggs serves to enhance the endemic state, that is, it increases the number of infected people.
		
The risk of a dengue outbreak can be reduced if an intervention is made to regulate the trade in second hand tires by setting appropriate limits on their storage time. In the ideal case, where the tires are processed immediately upon reaching the urban area $\tau_ {s} << \tau_ {d}$, an outbreak of dengue fever does not take place. In addition, removal of tires works as a control measure in rural areas by reducing the overall density of mosquito eggs and therefore in turn the number of dengue cases falls. 
		
Diapause of mosquito eggs and vertical transmission of the virus make distinct contributions to the dynamics of disease transmission. While diapause can facilitate the spread of mosquitoes, increasing the risk of their invasion of new areas, the combination of diapause with vertical transmission can simultaneously spread the virus and cause a dengue outbreak. It is this process, mediated through the movement of reservoirs of eggs, which gives the transportation of tires such a crucial role in the spread of the disease, but also opens the opportunity for control measures. When applied to the regulated movement and storage of tires in the specific context of Puerto Rico, our model shows that even an incomplete and spatially-limited intervention can substantially reduce the disease burden experienced by both rural and urban populations. Much greater control could be achieved were tires to be reliably stored in a dry place with a limit on storage time when kept in the open. These are effective and relatively low-cost actions. 
		
By estimating the number of secondary human infections in the urban area caused by each human infection in the rural area, we provide a means by which to evaluate the impact of both tire transportation and regulatory mechanisms. This could be used by policy makers to highlight the efficacy of interventions in a striking and accessible fashion which emphasizes the direct social implications. 
		
\section*{Acknowledgments}
Carlos Castillo-Chavez and Anuj Mubayi made helpful comments on an early version of the manuscript. We are also grateful for support from the SAL-MCMSC to participate in MTBI where part of this work was developed. We are grateful for support from CONACyT and SEP-PRODEP Grant DSA/103.5/15/7449 (Mexican agencies).
\nolinenumbers

\appendix


\section*{Appendix A:  Urban offspring reproduction number}
\label{Appendix A}
Considering the last one four equation of urban model and doing $\dot{M_{SU}}+\dot{M_{IU}}$ and $\dot{E_{SU}}+\dot{E_{IU}}$ it is obtained. 

\begin{align}\label{urban-vector}
\dot{M}	&= \kappa\omega E - \epsilon M,\nonumber\\
\dot{E}&=\phi M- (\pi + \omega)E+\frac{r\chi}{\theta}\psi\left(\frac{\tau_s}{\tau_d}\right)E_{R}. \tag{A.1}
\end{align}

We calculate the urban offspring number, using the method ~\cite{Brauer2008} we write the system(\ref{urban-vector}) as $\mathfrak{\dot{X}}=\mathfrak{F}-\mathfrak{V}$.

\begin{equation*}
{\mathfrak{\dot{x}}}=\left(
\begin{array}{cc}
\dot{M}\\
\\
\dot{ E}\\
\end{array}
\right),
\ \
{\mathfrak{F}}=\left(
\begin{array}{cc}
\kappa\omega E\\
\\
0\\
\end{array}
\right),
\ \
{\mathfrak{V}}=\left(
\begin{array}{cc}
\epsilon M\\
\\
(\pi + \omega)E-\phi M\left(1-\frac{E}{C}\right)-\frac{r\chi}{\theta}\psi\left(\frac{\tau_s}{\tau_d}\right)E_{R}\\
\end{array}
\right).
\end{equation*}
The Jacobian matrices $F$ and $V$, associated with $\mathfrak{F}$ and $\mathfrak{V}$ respectively, at the vector free equilibrium $M^{*}=0$, $E^{*}=0$ are.

\begin{equation*}
{F}=\left(
\begin{array}{cc}
0&\kappa\omega \\
\\
0&0\\
\end{array}
\right),
\ \
{V}=\left(
\begin{array}{cc}
\epsilon & 0\\
\\
-\phi & (\pi + \omega)\\
\end{array}
\right),
\ \
{V^{-1}}=\left(
\begin{array}{cc}
\frac{1}{\epsilon} & 0\\
\\
\frac{\phi}{\epsilon(\pi + \omega)} & \frac{1}{\pi + \omega}\\
\end{array}
\right),
\end{equation*}

\begin{equation*}
{K}=FV^{-1}=\left(
\begin{array}{cc}
\frac{\kappa\omega\phi}{\epsilon(\pi + \omega)} & \frac{\kappa\omega}{\pi + \omega}\\
\\
0 & 0\\
\end{array}
\right).
\end{equation*}
The eigenvalues of $K$ are $0$ and $\frac{\kappa\omega\phi}{\epsilon(\pi + \omega)}$, so the urban offspring reproduction number is given by:
\begin{equation*}
R_{M}^{u}={\kappa\omega\phi\over\epsilon(\pi+\omega)}
\end{equation*}

\section*{ Appendix B:   Vector demography}
\label{Appendix B}
Considering the last one four equation of rural model 
and doing $\dot{M_{SR}}+\dot{M_{IR}}$ and $\dot{E_{SR}}+\dot{E_{IR}}$ it is obtained. 

\begin{align}\label{demography}
\dot{M}	&= \kappa\omega E - \epsilon M,\nonumber\\
\dot{E}&= \phi M- (\pi + \omega +\frac{r}{\theta})E.\tag{B.1}
\end{align}

We calculate the basic offspring number of rural mosquitoes, using the method ~\cite{Brauer2008} we write the system(\ref{demography}) as $\mathfrak{\dot{X}}=\mathfrak{F}-\mathfrak{V}$.

\begin{equation*}
{\mathfrak{\dot{x}}}=\left(
\begin{array}{cc}
\dot{M}\\
\\
\dot{ E}\\
\end{array}
\right),
\ \
{\mathfrak{F}}=\left(
\begin{array}{cc}
\kappa\omega E\\
\\
0\\
\end{array}
\right),
\ \
{\mathfrak{V}}=\left(
\begin{array}{cc}
\epsilon M\\
\\
(\pi + \omega+\frac{r}{\theta})E-\phi M\left(1-\frac{E}{C}\right)\\
\end{array}
\right).
\end{equation*}
The Jacobian matrices $F$ and $V$, associated with $\mathfrak{F}$ and $\mathfrak{V}$ respectively, at the vector free equilibrium $M^{*}=0$, $E^{*}=0$ are.

\begin{equation*}
{F}=\left(
\begin{array}{cc}
0&\kappa\omega \\
\\
0&0\\
\end{array}
\right),
\ \
{V}=\left(
\begin{array}{cc}
\epsilon & 0\\
\\
-\phi & (\pi + \omega +\frac{r}{\theta})\\
\end{array}
\right),
\ \
{V^{-1}}=\left(
\begin{array}{cc}
\frac{1}{\epsilon} & 0\\
\\
\frac{\phi}{\epsilon(\pi + \omega+\frac{r}{\theta})} & \frac{1}{\pi + \omega + \frac{r}{\theta}}\\
\end{array}
\right),
\end{equation*}

\begin{equation*}
{K}=FV^{-1}=\left(
\begin{array}{cc}
\frac{\kappa\omega\phi}{\epsilon(\pi + \omega + \frac{r}{\theta})} & \frac{\kappa\omega}{\pi + \omega + \frac{r}{\theta}}\\
\\
0 & 0\\
\end{array}
\right).
\end{equation*}
The eigenvalues of $K$ are $0$ and $\frac{\kappa\omega\phi}{\epsilon(\pi + \omega + \frac{r}{\theta})}$, so the rural offspring reproduction number is given by:
\begin{equation*}
R_{M}^{r}={\kappa\omega\phi\over\epsilon(\pi+\omega +\frac{r}{\theta})}
\end{equation*}

The system has two stationary states $E^{*}=M^{*}=0$ and $M^{*}=\frac{C\kappa\omega}{\epsilon}\left(\frac{R_{M}^{r}-1}{R_{M}^{r}}\right)$, $E^{*}=C\left(\frac{R_{M}^{r}-1}{R_{M}^{r}}\right)$.

The linearizing around the trivial stationary solutions is done. For this, we will calculate the Jacobian matrix around the equilibrium point $(0,0)$.
\begin{equation*}
{\bf DF(0,0)}=\left(
\begin{array}{cc}
-\epsilon & \kappa\omega\\
\\
\phi &  -(\pi + \omega + \frac{r}{\theta})\\
\end{array}
\right).
\end{equation*}
We get the following characteristic polynomial,

\begin{equation*}\label{caracteristica}
\lambda^{2}+ (\epsilon+\pi+\omega+\frac{r}{\theta})\lambda + \epsilon(\pi+\omega+\frac{r}{\theta})(1-R_{M})=0
\end{equation*}
whose roots are of the shape
\begin{equation*}\label{raices}
\lambda_{\pm}=\frac{1}{2}(\gamma\pm\sqrt{\xi})
\end{equation*}
where $\gamma=-(\epsilon+\pi+\omega+\frac{r}{\theta})$, $\xi=\gamma^{2}-4\Xi$ and $\Xi=\epsilon(\pi+\omega+ \frac{r}{\theta})(1- R_{M})$.\\

\section*{ Appendix C: Basic reproduction number without vertical transmission in urban area}
\label{Appendix C}

In the following we calculate the urban reproductive number. The infected classes on urban model are: $I_{U}$, $E_{IU}$ and $M_{IU}$, so the matrices $\mathfrak{F}$ and $\mathfrak{V}$ take the following shape.

\begin{equation*}
\mathfrak{F}=\left( \begin{array}{c c c}
\vspace{.2cm}
\alpha\frac{S_{U}}{N_{U}}M_{IU}\\
\vspace{.2cm}
0\\
\vspace{.2cm}
\alpha \frac{I_{U}}{N_{U}}M_{SU}\\
\end{array}
\right)
\hspace{2cm}
\mathfrak{V}=\left( \begin{array}{c c c}
\vspace{.2cm}
(\eta+\gamma)I_{IU}\\
\vspace{.2cm}
(\pi+\omega)E_{IU}-\frac{r}{\theta}\chi\psi(\tau_{s}/\tau_{d})E_{IR}\\
\vspace{.2cm}
\epsilon M_{IU}-\kappa\omega E_{IU}\\
\end{array}
\right)
\end{equation*}
The Jacobian matrices are:

\begin{equation*}
F=\left( \begin{array}{c c c}
\vspace{.2cm}
0&0&\alpha\frac{S_{U}}{N_{U}}\\
\vspace{.2cm}
0&0&0\\
\vspace{.2cm}
\alpha \frac{M_{SU}}{N_{U}}&0&0\\
\end{array}
\right)
\hspace{.3cm}
V=\left( \begin{array}{c c c}
\vspace{.2cm}
(\eta+\gamma)&0&0\\
\vspace{.2cm}
0&(\pi+\omega)&0\\
\vspace{.2cm}
0&-\kappa\omega&\epsilon \\
\end{array}
\right)
\end{equation*}

\begin{equation*}
V^{-1}=\left( \begin{array}{c c c}
\vspace{.2cm}
\frac{1}{(\eta+\gamma)}&0&0\\
\vspace{.2cm}
0&\frac{1}{\pi+\omega}&0\\
\vspace{.2cm}
0&\frac{\kappa\omega}{\epsilon(\pi+\omega)}&\frac{1}{\epsilon} \\
\end{array}
\right)
\end{equation*}
Then we evaluated the Jacobian matrices at the disease free equilibrium $S_{U}=N_{U}, M_{SU}=M_{U},\  E_{SU}=E_{U},\ I_{U}=M_{IU}=E_{IU}=0$. Then we  found the eigenvalues of $K=FV^{-1}$ since the basic reproductive number is the spectral radius, then we need the maximum of the eigenvalues of $K$, this will be the basic reproductive number.
\begin{equation*}
K=\left( \begin{array}{c c c}
\vspace{.2cm}
0&\frac{\alpha\kappa\omega}{\epsilon(\pi+\omega)} & \frac{\alpha}{\epsilon}\\
\vspace{.2cm}
0&0& 0\\
\vspace{.2cm}
\frac{\alpha N}{(\eta+\gamma)M^{*}}&0&0\\
\end{array}
\right)
\end{equation*}

There are three eigenvalues, one of them is zero, the other is smaller, so the maximum is
\begin{eqnarray*}
	R_{0}^{u}&=&\sqrt{\frac{\alpha N}{(\mu+\gamma)M^{*}}\frac{\alpha}{\epsilon}}
\end{eqnarray*}

\section*{Appendix D: Basic reproduction number in rural area}
\label{Appendix D}
We calculated the basic reproductive number using the next generation matrix method $\mathfrak{\dot{X}}=\mathfrak{F}-\mathfrak{V}$ \cite{Brauer2008}. The infected classes on rural model are: $I_{R}$, $E_{IR}$ and $M_{IR}$, then, the information is separated into two matrices, the first one corresponds to new infection and the second one corresponds to disease progression, that is
\begin{equation*}
\mathfrak{\dot{X}}=\left(\begin{array}{c}
\vspace{.2cm}
\dot{I} \\
\vspace{.2cm}
\dot{E_{IR}}\\
\vspace{.2cm}
\dot{M_{IR}}\\
\end{array}
\right)
\end{equation*}

\begin{equation*}
\mathfrak{F}=\left( \begin{array}{c c c}
\vspace{.2cm}
\alpha\frac{S_{R}}{N_{R}}M_{IR}\\
\vspace{.2cm}
\nu\phi M_{IR}\left(1-\frac{E_{R}}{C_{r}}\right)\\
\vspace{.2cm}
\alpha \frac{I_{R}}{N_{R}}M_{SR}\\
\end{array}
\right)
\hspace{2cm}
\mathfrak{V}=\left( \begin{array}{c c c}
\vspace{.2cm}
(\eta+\gamma)I_{IR}\\
\vspace{.2cm}
(\pi+\omega+\frac{r}{\theta})E_{IR}\\
\vspace{.2cm}
\epsilon M_{IR}-\kappa\omega E_{IR}\\
\end{array}
\right)
\end{equation*}

The Jacobian matrices are:

\begin{equation*}
F=\left( \begin{array}{c c c}
\vspace{.2cm}
0&0&\alpha\frac{S_{R}}{N_{R}}\\
\vspace{.2cm}
0&-\frac{\nu\phi M_{IR}}{C_{r}}&\nu\phi\left(1-\frac{E_{R}}{C_{r}}\right)\\
\vspace{.2cm}
\alpha \frac{M_{SR}}{N_{R}}&0&0\\
\end{array}
\right)
\hspace{.3cm}
V=\left( \begin{array}{c c c}
\vspace{.2cm}
(\eta+\gamma)&0&0\\
\vspace{.2cm}
0&(\pi+\omega)+\frac{r}{\theta}&0\\
\vspace{.2cm}
0&-\kappa\omega&\epsilon \\
\end{array}
\right)
\end{equation*}

\begin{equation*}
V^{-1}=\left( \begin{array}{c c c}
\vspace{.2cm}
\frac{1}{(\eta+\gamma)}&0&0\\
\vspace{.2cm}
0&\frac{\theta}{\theta(\pi+\omega)+r}&0\\
\vspace{.2cm}
0&\frac{\theta\kappa\omega}{\epsilon(\theta(\pi+\omega)+r)}&\frac{1}{\epsilon} \\
\end{array}
\right)
\end{equation*}

We evaluated the Jacobian matrices at the disease free equilibrium $S_{R}=N_{R}$, $M_{SR}=M_{R}$, $E_{SR}=E_{R},\ I_{R}=M_{IR}=E_{IR}=0$. Then we  found the eigenvalues of $K=FV^{-1}$ since the basic reproductive number is the spectral radius. From this we need the maximum of the eigenvalues of $K$, which is the basic reproductive number.
\begin{equation*}
K=\left( \begin{array}{c c c}
\vspace{.2cm}
0&\frac{\alpha\theta\kappa\omega}{\epsilon(\theta(\pi+\omega)+r)} & \frac{\alpha}{\epsilon}\\
\vspace{.2cm}
0&\frac{\nu\phi\theta\kappa\omega}{\epsilon(\theta(\pi+\omega)+r)}\left(1-\frac{E}{C_{r}}\right)& \frac{\nu\phi}{\epsilon}\left(1-\frac{E}{C_{r}}\right)\\
\vspace{.2cm}
\frac{\alpha N}{M^{*}\eta+\gamma}&0&0\\
\end{array}
\right)
\end{equation*}

There are three eigenvalues, one of them is zero, the other is smaller, so the maximum is
{\small
\begin{eqnarray*}
		R_{0}^{r}&=&\frac{1}{2}\frac{\nu\phi\kappa\omega\theta}{\epsilon(\theta(\pi+\omega)+r)}\left(1-\frac{E_{R}^{*}}{C_{r}}\right)+\frac{1}{2}\sqrt{\left(\frac{\nu\phi\kappa\omega\theta}{\epsilon(\theta(\pi+\omega)+r)}\left(1-\frac{E_{R}^{*}}{C_{r}}\right)\right)^{2}+\frac{4\alpha }{\epsilon }\frac{\alpha N}{(\eta+\gamma)M^{*}}}
\end{eqnarray*}
}

\section*{ Appendix E: Basic reproduction number in urban area}
\label{Appendix E}

We calculate the urban reproductive number. The infected classes in the urban model are: $I_{U}$, $E_{IU}$ and $M_{IU}$, so the matrices $\mathfrak{F}$ and $\mathfrak{V}$ take the following shape:

\begin{equation*}
\mathfrak{F}=\left( \begin{array}{c c c}
\vspace{.2cm}
\alpha\frac{S_{U}}{N_{U}}M_{IU}\\
\vspace{.2cm}
\nu\phi M_{IU}\left(1-\frac{E_{U}}{C_{u}}\right)\\
\vspace{.2cm}
\alpha \frac{I_{U}}{N_{U}}M_{SU}\\
\end{array}
\right)
\hspace{2cm}
\mathfrak{V}=\left( \begin{array}{c c c}
\vspace{.2cm}
(\eta+\gamma)I_{IU}\\
\vspace{.2cm}
(\pi+\omega)E_{IU}-\frac{r}{\theta}\chi\psi(\tau_{s}/\tau_{d})E_{IR}\\
\vspace{.2cm}
\epsilon M_{IU}-\kappa\omega E_{IU}\\
\end{array}
\right)
\end{equation*}

The Jacobian matrices are:

\begin{equation*}
F=\left( \begin{array}{c c c}
\vspace{.2cm}
0&0&\alpha\frac{S_{U}}{N_{U}}\\
\vspace{.2cm}
0&-\frac{\nu\phi M_{IU}}{C_{u}}&\nu\phi\left(1-\frac{E_{U}}{C_{u}}\right)\\
\vspace{.2cm}
\alpha \frac{M_{SU}}{N_{U}}&0&0\\
\end{array}
\right)
\hspace{.3cm}
V=\left( \begin{array}{c c c}
\vspace{.2cm}
(\eta+\gamma)&0&0\\
\vspace{.2cm}
0&(\pi+\omega)&0\\
\vspace{.2cm}
0&-\kappa\omega&\epsilon \\
\end{array}
\right)
\end{equation*}

\begin{equation*}
V^{-1}=\left( \begin{array}{c c c}
\vspace{.2cm}
\frac{1}{(\eta+\gamma)}&0&0\\
\vspace{.2cm}
0&\frac{1}{\pi+\omega}&0\\
\vspace{.2cm}
0&\frac{\kappa\omega}{\epsilon(\pi+\omega)}&\frac{1}{\epsilon} \\
\end{array}
\right)
\end{equation*}
Then we evaluated the Jacobian matrices at the disease free equilibrium $S_{U}=N_{U}, M_{SU}=M_{U},\  E_{SU}=E_{U},\ I_{U}=M_{IU}=E_{IU}=0$. We found the eigenvalues of $K=FV^{-1}$ since the basic reproductive number is the spectral radius. The maximum of the eigenvalues of $K$ is the basic reproductive number.
\begin{equation*}
K=\left( \begin{array}{c c c}
\vspace{.2cm}
0&\frac{\alpha\kappa\omega}{\epsilon(\pi+\omega)} & \frac{\alpha}{\epsilon}\\
\vspace{.2cm}
0&\frac{\nu\phi\kappa\omega}{\epsilon(\pi+\omega)}\left(1-\frac{E_{U}}{C_{u}}\right)& \frac{\nu\phi}{\epsilon}\left(1-\frac{E_{U}}{C_{u}}\right)\\
\vspace{.2cm}
\frac{\alpha N}{(\eta+\gamma)M^{*}}&0&0\\
\end{array}
\right)
\end{equation*}

There are three eigenvalues, one of them is zero, the other is smaller, so the maximum is

\begin{eqnarray*}
	R_{0}^{u}&=& \frac{1}{2}\frac{\nu\phi\kappa\omega}{\epsilon(\pi+\omega)}\left(1-\frac{E_{U}^{*}}{C_{u}}\right)+\frac{1}{2}\sqrt{\left(\frac{\nu\phi\kappa\omega}{\epsilon(\pi+\omega)}\left(1-\frac{E_{U}^{*}}{C_{u}}\right)\right)^{2}+\frac{4\alpha }{\epsilon }\frac{\alpha N}{(\eta+\gamma) M^{*}}}
\end{eqnarray*}

The basic reproductive number  of the complete model is the maximum of the two reproductive numbers, \textit{rural reproduction number} and \textit{urban reproduction number}.

\begin{eqnarray*}
	R_{0}&=&\max{\{R_{0}^{r},\  R_{0}^{u}\}}
\end{eqnarray*}

\section*{ Appendix F:   Number of transmissions from rural to urban area}
\label{Appendix F}
To find transmission from rural to urban reproduction number, we following the same idea to calculate the above  basic reproduction numbers, so we want to know how many infections could cause an individual of the rural population in the urban population by the movement of infected tires, for that we are consider that in rural population the disease is endemic. The infected classes on full model are: $I_{R},\ E_{IR},\ M_{IR},\ I_{U},\ E_{IU},\ M_{IU}$.  The information is separated into two matrices, the first one corresponds to  new infection $\mathfrak{F}$ and the second one corresponds to disease progression $\mathfrak{V}$, that is

\begin{equation*}
\mathfrak{F}=\left( \begin{array}{c c c}
\vspace{.2cm}
\alpha\frac{S_{U}}{N_{U}}M_{IU}\\
\vspace{.2cm}
\nu\phi M_{IU}\left(1-\frac{E_{U}}{C_{u}}\right)\\
\vspace{.2cm}
\alpha \frac{I_{U}}{N_{U}}M_{SU}\\
\vspace{.2cm}
\alpha\frac{S_{R}}{M_{R}}M_{IR}\\
\vspace{.2cm}
\nu\phi M_{IR}\left(1-\frac{E_{R}}{C_{r}}\right)\\
\vspace{.2cm}
\alpha \frac{I_{R}}{N_{R}}M_{SR}\\
\end{array}
\right)
\hspace{2cm}
\mathfrak{V}=\left( \begin{array}{c c c}
\vspace{.2cm}
(\eta+\gamma)I_{IU}\\
\vspace{.2cm}
(\pi+\omega)E_{IU}-\frac{r}{\theta}\chi\psi(\tau_{s}/\tau_{d})E_{IR}\\
\vspace{.2cm}
\epsilon M_{IU}-\kappa\omega E_{IU}\\
\vspace{.2cm}
(\eta+\gamma)I_{IR}\\
\vspace{.2cm}
(\pi+\omega)E_{IR}+\frac{r}{\theta}E_{IR}\\
\vspace{.2cm}
\epsilon M_{IR}-\kappa\omega E_{IR}\\
\end{array}
\right)
\end{equation*}
The Jacobian matrices are:

\begin{equation*}
F=\left( \begin{array}{c c c c c c}
\vspace{.2cm}
0&\alpha\frac{S_{U}}{N_{U}}&0&0&0&0\\
\vspace{.2cm}
\alpha \frac{M_{SU}}{N_{U}}&0&0&0&0&0\\
\vspace{.2cm}
\vspace{.2cm}
0&\nu\phi\left(1-\frac{E_{U}}{C_{u}}\right)&-\frac{\nu\phi M_{IU}}{C_{u}}&0&0&0\\
\vspace{.2cm}
0&0&0&0&\alpha\frac{S_{R}}{N_{R}}&0\\
\vspace{.2cm}
0&0&0&\alpha \frac{M_{SR}}{N_{R}}&0&0\\
0&0& 0&0&\nu\phi\left(1-\frac{E_{R}}{C_{r}}\right)&-\frac{\nu\phi M_{IR}}{C_{r}}\\

\end{array}
\right)
\end{equation*}

\begin{equation*}
V=\left( \begin{array}{c c c c c c}
\vspace{.2cm}
(\eta+\gamma)&0&0&0&0&0\\
\vspace{.2cm}
0&\epsilon&-\kappa\omega&0&0&0\\
\vspace{.2cm}
0&0&(\pi+\omega)&0&0&-\frac{r}{\theta}\chi\psi(\tau_{s}/\tau_{d}) \\
\vspace{.2cm}
0&0&0&(\eta+\gamma)&0&0\\
\vspace{.2cm}
0&0&0&0&\epsilon&-\kappa\omega\\
\vspace{.2cm}
0&0&0&0&0&(\pi+\omega+ \frac{r}{\theta}) \\
\end{array}
\right)
\end{equation*}

\begin{equation*}
V^{-1}=\left( \begin{array}{c c c c c c}
\vspace{.2cm}
\frac{1}{(\eta+\gamma)}&0&0&0&0&0\\
\vspace{.2cm}
0&\frac{1}{\epsilon}&\frac{\kappa\omega}{\epsilon(\pi+\omega)}&0&0&\frac{\kappa\omega r\chi}{\epsilon(\pi+\omega)(\theta(\pi+\omega)+r)}\psi\left(\frac{\tau_s}{\tau_d}\right)\\
\vspace{.2cm}
0&0&\frac{1}{(\pi+\omega)}&0&0&\frac{ r\chi}{(\pi+\omega)(\theta(\pi+\omega)+r)}\psi\left(\frac{\tau_s}{\tau_d}\right) \\
\vspace{.2cm}
0&0&0&\frac{1}{(\eta+\gamma)}&0&0\\
\vspace{.2cm}
0&0&0&0&\frac{1}{\epsilon}&\frac{\kappa\omega\theta}{\epsilon(\theta(\pi+\omega)+r)}\\
\vspace{.2cm}
0&0&0&0&0&\frac{\theta}{\theta(\pi+\omega+r)} \\
\end{array}
\right)
\end{equation*} 
Then we evaluated the Jacobian matrices at the disease free equilibrium. Then we  found $K=FV^{-1}$. To get the \textit{number of transmissions from rural to urban area} we obtain $K^{3}$  with this matrix in the column if infectious rural population we get the following.

\begin{eqnarray*}
	R_{r\rightarrow u}&=&\frac{\alpha\kappa\omega r\chi }{\epsilon(\omega+\pi)(\theta(\omega+\pi)+r)}\psi\left(\frac{\tau_{s}}{\tau_{d}}\right)\frac{\nu\phi}{\epsilon}\frac{\alpha}{(\eta+\gamma)}
\end{eqnarray*}

\section*{Appendix G: Summary of reproduction number }
\label{Appendix G}
\begin{itemize}
	\item Urban offspring reproduction number.\\
	\begin{equation*}
	R_{M}^{u}={\kappa\omega\phi\over\epsilon(\pi+\omega)}
	\end{equation*}
	\item Rural offspring reproduction number.\\
	\begin{equation*}
	R_{M}^{r}={\kappa\omega\phi\over\epsilon(\pi+\omega +\frac{r}{\theta})}
	\end{equation*}
	\item Basic reproduction number without vertical transmission in urban area.\\
	\begin{eqnarray*}
		R_{0}^{u}&=&\sqrt{\frac{\alpha N}{(\mu+\gamma)M^{*}}\frac{\alpha}{\epsilon}}
	\end{eqnarray*}
	\item Basic reproduction number in urban area.\\
	\begin{eqnarray*}
		R_{0}^{u}&=& \frac{1}{2}\frac{\nu\phi\kappa\omega}{\epsilon(\pi+\omega)}\left(1-\frac{E_{U}^{*}}{C_{u}}\right)+\frac{1}{2}\sqrt{\left(\frac{\nu\phi\kappa\omega}{\epsilon(\pi+\omega)}\left(1-\frac{E_{U}^{*}}{C_{u}}\right)\right)^{2}+\frac{4\alpha }{\epsilon }\frac{\alpha N}{(\eta+\gamma) M^{*}}}
	\end{eqnarray*}
	\item Basic reproduction number in rural area.\\
	{\small
		\begin{eqnarray*}
			R_{0}^{r}&=&\frac{1}{2}\frac{\nu\phi\kappa\omega\theta}{\epsilon(\theta(\pi+\omega)+r)}\left(1-\frac{E_{R}^{*}}{C_{r}}\right)+\frac{1}{2}\sqrt{\left(\frac{\nu\phi\kappa\omega\theta}{\epsilon(\theta(\pi+\omega)+r)}\left(1-\frac{E_{R}^{*}}{C_{r}}\right)\right)^{2}+\frac{4\alpha }{\epsilon }\frac{\alpha N}{(\eta+\gamma)M^{*}}}
		\end{eqnarray*}}
		\item Number of transmissions from rural to urban area.\\
		\begin{eqnarray*}
			R_{r\rightarrow u}&=&\frac{\alpha\kappa\omega r\chi }{\epsilon(\omega+\pi)(\theta(\omega+\pi)+r)}\psi\left(\frac{\tau_{s}}{\tau_{d}}\right)\frac{\nu\phi}{\epsilon}\frac{\alpha}{(\eta+\gamma)}
		\end{eqnarray*}
		
	\end{itemize}

 \bibliographystyle{elsarticle-harv} 
 \bibliography{biblio-reservoir,books-reservoirs}





\end{document}